\documentclass[fleqn,usenatbib]{mnras}
\usepackage{color,amsmath,amssymb,latexsym,txfonts}
\usepackage{graphicx,subfigure,threeparttable}
\newcounter{Note}[enumi]
\renewcommand{\theNote}{\alph{Note}}
\newcommand{\note}[1]{\refstepcounter{Note}$ ^{\text{\theNote}} $\label{#1}}
\newcommand{\refn}[1]{$ ^{\text{\ref{#1}}} $}
\begin{document}
\title[Constraints on DSA]{Global Constraints on Diffusive Particle Acceleration by Strong Nonrelativistic Shocks}
\author[Y. Zhang and S. Liu]{
Yiran Zhang$ ^{1,2} $\thanks{E-mail: zhangyr@pmo.ac.cn}
and Siming Liu$ ^{1,2} $\thanks{E-mail: liusm@pmo.ac.cn}
\\
$ ^1 $Key Laboratory of Dark Matter and Space Astronomy, Purple Mountain Observatory, Chinese Academy of Sciences, Nanjing 210034, China\\
$ ^2 $School of Astronomy and Space Science, University of Science and Technology of China, Hefei 230026, China
}
\maketitle
\begin{abstract}
Estimating the cosmic-ray acceleration efficiency $ \epsilon $ in supernova remnants (SNRs) through observations is a challenging task in general. Based on the Rankine-Hugoniot shock conditions, we find an anticorrelation between $ \epsilon $ and the power-law spectral index $ \alpha $ of relativistic particle distribution produced via diffusive particle acceleration by nonrelativistic shocks, implying more efficient acceleration in older SNRs with harder radio spectra. Then $ \epsilon $ may be estimated from some hard radio spectral index measurements. Assuming the particle distribution in downstream of strong shocks to be a nonrelativistic Maxwellian plus a relativistic power law with a high-energy cutoff, we also find that the injection rate for relativistic particles $ \eta $ needs to $ \gtrsim 10^{-6} $ for a prominent decrease of the adiabatic index in SNRs, which implies higher compression ratio and lower values of $ \alpha $. This threshold of $ \eta $ increases with the shock speed $ u_1 $, which may explain the relatively harder radio spectra of older SNRs with lower $ u_1 $. We show that $ \eta $ and/or the relativistic cutoff momentum $ p_\text{m} $ need to be low for old SNRs, and expect a gradual increase of $ \epsilon $ as SNR evolves with gradually decreasing $ \eta $ and $ p_\text{m} $.
\end{abstract}
\section{Introduction}\label{s1}
Diffusive shock acceleration (DSA) is considered the primary mechanism of astrophysical particle acceleration. This mechanism can easily achieve a power-law distribution of accelerated particles in the momentum space, and is widely used to explain nonthermal distributions occurring in varieties of astrophysical environments, such as in solar flares and supernova remnants (SNRs). In this paper, we focus on global characteristics of strong nonrelativistic shocks without considering details of physical processes near the shock front.

In the test particle limit, the classical DSA theory \citep{1983RPPh...46..973D} shows that planar shocks produce a power-law steady-state distribution of energetic particles (with a speed much higher than the shock speed) in the momentum space in the downstream with the spectral index
\begin{equation}
\alpha =\frac{3r}{r-1},\label{e38}
\end{equation}
where $ r=u_1/u_2 $ is the shock compression ratio, $ u $ is the nonrelativistic fluid speed in the shock frame, subscripts 1 and 2 represent upstream and downstream, respectively. For a strong shock with nonrelativistic downstream gas, the adiabatic index $ \gamma _2=5/3 $, $ r=4 $, thus $ \alpha =4 $. If the downstream gas becomes relativistic with $ \gamma _2<5/3 $, $ r $ can be larger than 4 leading to an even harder particle distribution, which may contain enough energy to reduce $ \gamma _2 $ significantly. Therefore effects of accelerated particles on the global characteristics of the DSA may be obtained self-consistently without considering details of the particle acceleration processes near the shock front \citep{1981ApJ...244..711E,1984ApJ...286..691E,2005ApJ...632..920E,2010APh....33..307C,2010ApJ...708..965Z}.

In the general self-consistent DSA theory \citep{1985PhRvL..55.2735E,1999ApJ...526..385B}, the shock structure is modified by accelerated particles streaming away from the shock front in the upstream leading to a precursor, where the flow speed gradually approaches $ u_1 $ away from the shock front. The details of these nonlinear effects depend on the poorly understood processes of particle injection to DSA at low energies and the energy dependence of the diffusion coefficient near the shock front. The steady-state nonthermal distribution is then modified to a concave spectral shape, because particles with higher energies usually diffuse farther and experience larger effective compression ratio than particles with lower energies leading to a spectral hardening toward high energies. The high-energy spectral index should approach $ \alpha $ given by Equation (\ref{e38}) as the corresponding energetic particles may experience the whole shock structure. The dependence of $ \alpha $ on the acceleration efficiency does not change significantly with properties of the shock precursor and is the focus of the current investigation.

The cosmic-ray (CR) acceleration efficiency $ \epsilon $ is an essential parameter in the hypothesis that Galactic CRs (below the spectral ``knee'') originate mainly from SNRs, and an $ \epsilon $ of a few tens of percent is needed to match the measured CR energy density \citep{2012SSRv..173..369H}. However, the correlation between $ \epsilon $ and $ \alpha $ has not been explored since it is difficult to measure the spectral index of high-energy particle distribution, and GeV electrons producing the observed radio emission of SNRs are usually considered not energetic enough to experience the whole shock structure. Then the electron spectral index inferred from radio observations should be considered as an upper limit for $ \alpha $, which can lead to a lower limit for $ \epsilon $. Moreover, $ \gamma $-ray observations of SNRs reveal a spectral softening toward high energies in opposite to predictions of some nonlinear DSA models.

More interestingly, recent observations of RX J1713-3946 imply an energy-independent diffusion coefficient in the shock upstream \citep{2018A&A...612A...6H}, which indicates that the particle diffusion may be dominated by turbulent convection \citep[turbulent diffusion:][]{1988SvAL...14..255P,1993PhyU...36.1020B}. \cite{2017ApJ...844L...3Z} showed that a weak dependence of the diffusion coefficient on the particle energy may also explain anomalous distributions of near-earth CR spectra by considering a time-dependent particle acceleration scenario in SNRs. The energy-independent diffusion implies a fast transition between the low-energy thermal and the high-energy power-law distribution, so that the radio spectral observations can be used to obtain $ \alpha $, and we may introduce an injection rate to characterize the injection process. Although the classical DSA has succeeded in explaining many observational characteristics of SNRs \citep{2012SSRv..173..369H}, the gradual hardening of radio spectrum with age and the convex $ \gamma $-ray spectrum remain to be addressed \citep{2008ARA&A..46...89R,2013A&A...553A..34D}. Further investigation of the DSA is warranted.

In this work, we shall assume that the relativistic CR distribution reaches the steady-state slope given by Equation (\ref{e38}), i.e., particles in the downstream have a Maxwell distribution at nonrelativistic energies and power-law one with a high-energy cutoff at relativistic energies. For nonlinear DSAs, the actual distribution of relativistic particles should be softer and the $ \alpha $ in this paper should be considered as a lower limit. The injection rate for relativistic particles then characterizes the injection process. The high-energy cutoff can partially take into account the time-dependent acceleration effect or nonlinear effect of magnetic field amplification by CRs streaming away from the shock front \citep{2004MNRAS.353..550B}.

The outline of this paper is as follows. In Section \ref{s2}, we study the dependence of $ \epsilon $ on $ \alpha $. In Section \ref{s3}, we study the detailed self-consistent solutions with the assumed particle distribution mentioned above. In Section \ref{s4}, we give some discussions.
\section{CR Acceleration Efficiency}\label{s2}
Shock conditions follow the conservation of particle numbers, energies, and momenta across the shock front in the shock frame, where the shock front is at rest. For an ideal fluid we have \citep{1948PhRv...74..328T}
\begin{align}
\rho _1\varGamma _1u_1&=\rho _2\varGamma _2u_2,
\label{e1}\\
H_1\varGamma _{1}^{2}u_1&=H_2\varGamma _{2}^{2}u_2,\label{e2}\\
P_1+H_1\left( \frac{\varGamma _1u_1}{c} \right) ^2&=P_2+H_2\left( \frac{\varGamma _2u_2}{c} \right) ^2,\label{e3}
\end{align}
where $ \varGamma $ is the Lorentz factor corresponding to the fluid speed $ u $, $ c $ is the speed of light, $ H=\rho c^2+U+P $ is the relativistic enthalpy density, and
\begin{equation}
\rho =\int{mf\text{d}^3p},\ U=\int{\varepsilon f\text{d}^3p},\ P=\int{\frac{pv}{3}f\text{d}^3p}\label{e4}
\end{equation}
are the rest mass density, the internal energy density, the gas pressure, respectively, and $ m $ is the rest mass of particle. These thermodynamic functions are defined in the fluid co-moving frame, as the particle's kinetic energy $ \varepsilon $, (magnitude of) momentum $ p $, speed $ v $ and phase space distribution function $ f $ involved in Equations (\ref{e4}) are all measured in this frame, and $ f $ must be approximately isotropic (within the scope of hydrodynamics). The relativistic hydrodynamics reduces to the classical nonrelativistic formulas only if $ u\ll c $ and $ U\ll \rho c^2 $. From Equations (\ref{e1}) $ \sim $ (\ref{e3}), we have
\begin{align}
\left( \frac{u_1}{c} \right) ^2&=\frac{\left( \rho _2c^2+U_2+P_1 \right) \left( P_1-P_2 \right)}{\left( \rho _1c^2+U_1+P_2 \right) \left[ \left( \rho _1-\rho _2 \right) c^2+U_1-U_2 \right]},\label{e5}\\
\left( \frac{u_2}{c} \right) ^2&=\frac{\left( \rho _1c^2+U_1+P_2 \right) \left( P_1-P_2 \right)}{\left( \rho _2c^2+U_2+P_1 \right) \left[ \left( \rho _1-\rho _2 \right) c^2+U_1-U_2 \right]}.\label{e6}
\end{align}
For the case of nonrelativistic shocks $ u_1\ll c $, $ U/\left( \rho c^2\right) \sim O\left( u/c\right)^2 $, one may use the classical nonrelativistic hydrodynamics, and Equations (\ref{e1}), (\ref{e5}) and (\ref{e6}) then reduce to the Rankine-Hugoniot conditions:
\begin{align}
\frac{\rho _1}{\rho _2}&=\frac{2U_1+P_1+P_2}{2U_2+P_1+P_2},\label{e7}\\
u_1^2&=\frac{P_1-P_2}{\left( \rho _1/\rho _2-1 \right) \rho _1},\label{e8}\\
u_2^2&=\frac{P_1-P_2}{\left( 1-\rho _2/\rho _1 \right) \rho _2}.\label{e9}
\end{align}

We are interested in the global characteristics of the DSA by a one-dimensional planar shock. Then the subscript 1 represents far upstream, where the presence of CR precursor can be neglected, and the subscript 2 represents the homogeneous downstream. The \emph{overall} shock compression ratio $ r=\rho _2/\rho _1=u_1/u_2 $ can be derived from Equations (\ref{e7}) and (\ref{e8}):
\begin{equation}
r=\frac{\frac{1}{\gamma _1}+M^2+\sqrt{\left( \frac{1}{\gamma _1}+M^2 \right) ^2+\left( \frac{1}{\gamma _2^2}-1 \right) \left( \frac{2}{\gamma _1-1}+M^2 \right) M^2}}{\left( 1-\frac{1}{\gamma _2} \right) \left( \frac{2}{\gamma _1-1}+M^2 \right)},\label{e30}
\end{equation}
where
\begin{equation}
\gamma =1+\frac{P}{U},\ M=\sqrt{\frac{\rho _1}{\gamma _1P_1}}u_1,\label{e31}
\end{equation}
are the adiabatic index and the shock Mach number, respectively. As can be seen the compression ratio not only depends on the Mach number and the adiabatic index in the upstream, but also depends on the adiabatic index in the downstream, which is determined by the acceleration process. In general, particle escape is not necessary for efficient particle acceleration, which contradicts results of \cite{2010ApJ...722.1727V}, where the effect of accelerated particles on the dynamics of the subshock is ignored.

If one adopts a free escaping boundary condition of accelerated particles in the upstream, there will be a net flux of accelerated particles in the steady-state, contributions of this flux to conservation Equations (\ref{e1}) $ \sim $ (\ref{e3}) need to be incorporated properly to study the effects of CR escape on the shock compression ratio. Radiative loss and escape of CRs near the shock front can be added to the energy conservation directly \citep{2010ApJ...722.1727V}. For strong shocks with $ M\rightarrow \infty $, the compression ratio only depends on $ \gamma _2 $ with $ r = \left( \gamma _2+1\right) /\left( \gamma _2-1\right) $. For $ \gamma _1=\gamma _2 $, Equation (\ref{e30}) reduces to the standard formula:
\begin{figure*}
	\includegraphics[width=1\textwidth]{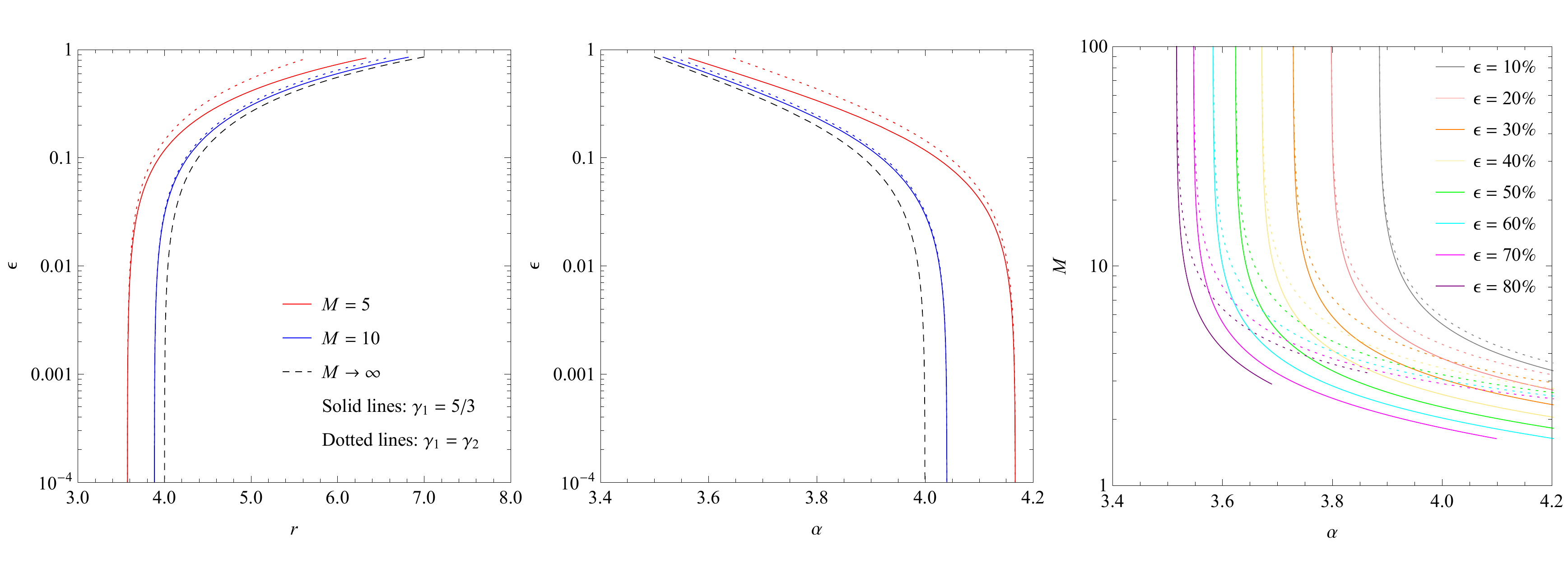}
	\caption{Dependence of $ \epsilon $ on $ r $ (left) and $ \alpha $ (middle), and contours of $ \epsilon $ in the parameter space $ \left( \alpha ,M\right) $ (right) as derived from Equations (\ref{e34}) $ \sim $ (\ref{e36}).\label{f1}}
\end{figure*}
\begin{equation}
r=\frac{\left( \gamma +1 \right) M^2}{\left( \gamma -1 \right) M^2+2}.
\end{equation}

For a DSA shock, we split the downstream pressure $ P_2=P_\text{nr}+P_\text{cr} $, where subscripts nr and cr indicate the nonrelativistic and the relativistic composition (i.e., the CR composition), respectively. Clearly $ 4/3<\gamma _2<5/3 $ as $ U_{\text{nr}}=3P_{\text{nr}}/2 $ and $ U_{\text{cr}}=3P_{\text{cr}} $, thus
\begin{equation}
\frac{P_{\text{nr}}}{P_{\text{cr}}}=2\frac{3\gamma _2-4}{5-3\gamma _2}.\label{e37}
\end{equation}
We define the CR acceleration efficiency as the ratio of the downstream CR pressure to the upstream total momentum flux density, then Equations (\ref{e7}), (\ref{e8}), (\ref{e31}) and (\ref{e37}) give
\begin{align}
\epsilon =\frac{P_{\text{cr}}}{P_1+\rho _1u_{1}^{2}}&=\frac{\gamma _1M^2}{1+\gamma _1M^2}\frac{2}{1+\gamma _2}\frac{P_{\text{cr}}}{P_2}\notag\\
&=\frac{\gamma _1M^2}{1+\gamma _1M^2}\frac{2\left( 5-3\gamma _2 \right)}{3\left( \gamma _2^2-1 \right)},\label{e34}
\end{align}
where
\begin{equation}
\gamma _2=\frac{\left( \frac{2}{\gamma _1-1}+M^2 \right) r^2-M^2}{\left( \frac{2}{\gamma _1-1}+M^2 \right) r^2-2\left( \frac{1}{\gamma _1}+M^2 \right) r+M^2}\label{e35}
\end{equation}
is derived from Equation (\ref{e30}). Obviously $ \gamma _2>4/3 $ gives an upper bound of $ \epsilon $:
\begin{equation}
\epsilon <\frac{6}{7}\frac{\gamma _1M^2}{1+\gamma _1M^2}<\frac{6}{7}.\label{e39}
\end{equation}
The case $ \gamma _1<\gamma _2 $ should also be excluded because shocks are expected to produce more relativistic particles in the downstream.

As indicated in Section \ref{s1}, one may use Equation (\ref{e38}) to replace $ r $ in Equation (\ref{e34}) with $ \alpha $ to obtain correlations between $ \epsilon $ and $ \alpha $. When $ \epsilon $ is high (say greater than 10\%), the nonlinear effects imply that the spectral index of accelerated particles should be larger than $ \alpha $. In general, the results depend on the adiabatic index in the upstream $ \gamma _1 $. For $ M\rightarrow \infty $,
\begin{equation}
\epsilon =\frac{3\left( 4-\alpha \right)}{\alpha \left( \alpha -3 \right)}.\label{e33}
\end{equation}
For $ \gamma _1=\gamma _2 $,
\begin{equation}
\epsilon =\frac{3M^2\left[ 2\left( M^2-1 \right) \alpha -3M^2 \right] \left[ \left( M^2-1 \right) \alpha -4M^2 \right]}{\left( M^2-1 \right) ^2\alpha \left( 3-2\alpha \right) \left[ \left( M^2-1 \right) \alpha -3M^2 \right]}.\label{e36}
\end{equation}
Dependence of $ \epsilon $ on $ r $ and $ \alpha $ are shown in the left and the middle panel of Figure \ref{f1}, respectively, and the right panel shows contours of $ \epsilon $ in the parameter space $ \left( \alpha ,M\right) $. For high values of $ \epsilon $, $ \alpha $ should be treated as a lower limit to spectral index of accelerated particles. As can be seen, $ \epsilon $ increases with the decrease of $ \alpha $ for a given $ M $, implying that older SNRs, which usually have harder radio spectra, are inclined to have more efficient acceleration. This is consistent with our two-stage acceleration model for anomalous distribution of Galactic CRs \citep{2017ApJ...844L...3Z}. Although it is generally accepted that acceleration of highest-energy CRs occurs in the early stage of SNR evolution with higher shock speed \citep{2004MNRAS.353..550B}, the overall energy density of CRs is dominated by GeV CRs for the soft CR spectrum. The two-stage acceleration model predicts that most of the GeV CRs are accelerated in old SNRs interacting with molecular clouds. For a given $ \alpha $ or $ r $, $ \epsilon $ also increases with the decrease of $ M $, implying that Equation (\ref{e33}) is a lower limit to $ \epsilon $. In particular, we note that $ \alpha <3.88 $ is a sufficient condition for $ \epsilon >10\% $.

One should note that several processes may make the evaluation of $ \epsilon $ with $ \alpha $ ambiguous. First of all, radiative loss and escape cooling can enhance the compression ratio, leading to a lower value of $ \alpha $. Moreover, acceleration of pre-existing CRs in the upstream, contributions to radio emission via the bremsstrahlung, Coulomb collision energy loss of sub-GeV particles and stochastic particle acceleration in the downstream may also lead to a harder radio spectrum \citep{1999ApJ...511..798C}. The CR acceleration efficiency obtained above therefore may be an overestimate. On the other hand, if the particle distribution has not reached the steady state or nonlinear DSA plays a significant role, the GeV electron distribution obtained from radio observations should be softer than that given by Equation (\ref{e38}), we expect even more efficient acceleration than derived above, therefore Equation (\ref{e33}) still works as a lower limit in this case. Clarification of these processes are necessary to have an accurate measurement of $ \epsilon $ for individual SNRs.
\begin{figure*}
	\includegraphics[width=1\textwidth]{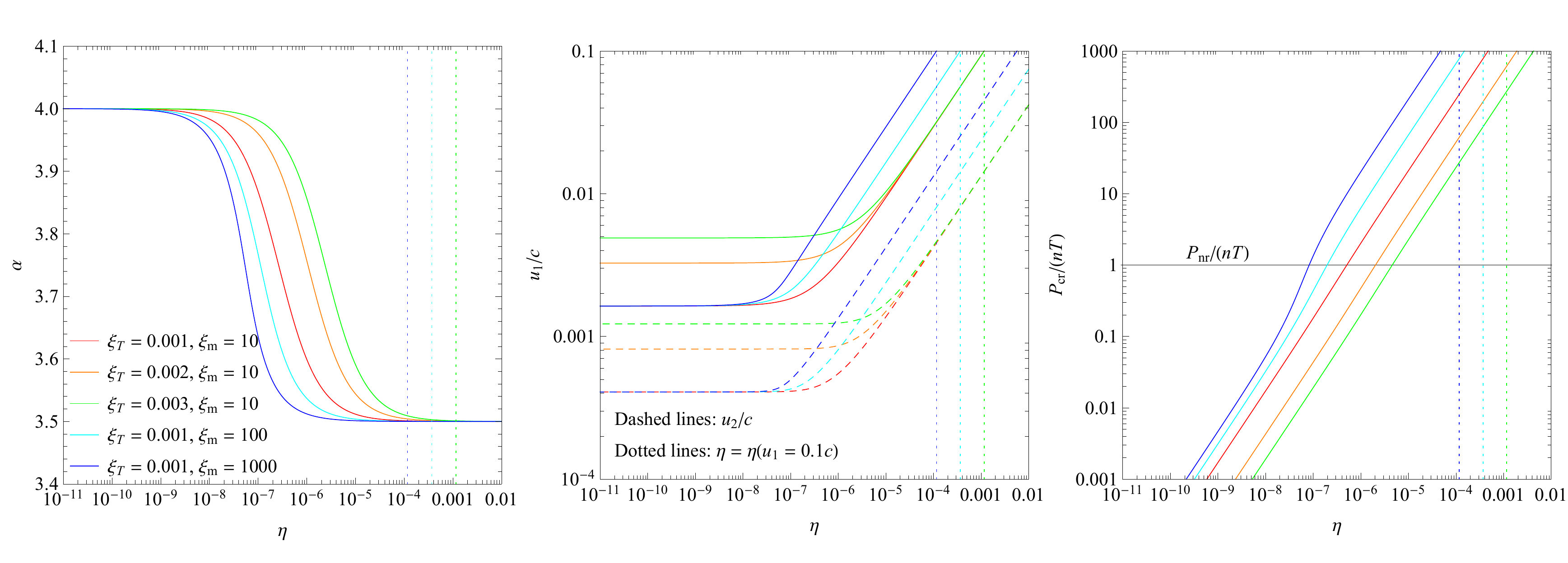}
	\caption{Dependence of $ \alpha $ (left), $ u $ (middle) and $ P $ (right) on $ \eta $ with $ \xi _T $ and $ \xi _\text{m} $ indicated in the left panel as derived from Equations (\ref{e8}), (\ref{e31}), (\ref{e11}) $ \sim $ (\ref{e10}). Each solution shown in this figure should be reasonable for $ u_1<0.1c $.\label{f2}}
\end{figure*}
\section{Constraint on The Injection Process}\label{s3}
Injection process of nonlinear DSA may be constrained with proper assumptions of the particle distribution in the downstream. Nonlinear DSA theories usually give a concaved spectrum of accelerated particles with the high-energy particle distribution approaching a power law with the index given by Equation (\ref{e38}) (if there are particles that can experience the whole shock structure), we however assume that dramatic concave spectrum occurs only when $ p<mc $. The injection rate of relativistic particles investigated below therefore should be considered as a lower limit. Although the number of relativistic particles can be neglected compared with the nonrelativistic one, the pressure carried by relativistic particles controls the adiabatic index and the acceleration of nonrelativistic particles does not play an important role in the overall shock dynamics. Therefore, we assume the downstream particle distribution to be Maxwellian at nonrelativistic energies and a power law with high-energy cutoff at relativistic energies. Considering Equations (\ref{e4}), we evaluate the equations of state (EOS):
\begin{align}
U_{\text{nr}}&=\frac{3}{2}nT=\frac{3}{4}nmc^2\xi _T^2=\frac{3}{2}P_{\text{nr}},\label{e11}\\
U_{\text{cr}}&\approx \eta n\frac{\int_{mc}^{p_{\text{m}}}{p^{-\alpha}cp4\pi p^2\text{d}p}}{\int_{mc}^{p_{\text{m}}}{p^{-\alpha}4\pi p^2\text{d}p}}=\eta nmc^2\frac{3-\alpha}{4-\alpha}\frac{\xi _{\text{m}}^{4-\alpha}-1}{\xi _{\text{m}}^{3-\alpha}-1}=3P_{\text{cr}},\label{e12}
\end{align}
where $ n\approx \rho _2/m $ and $ T $ are the particle number density and the thermal energy of the Maxwell distribution, respectively, $ \eta \approx n_{\text{cr}}/n $ is the injection rate for relativistic particles, $ p_{\text{m}} $ is the maximum momentum of the high-energy tail, and
\begin{equation}
\xi _T=\frac{\sqrt{2mT}}{mc},\ \xi _{\text{m}}=\frac{p_{\text{m}}}{mc}.\label{e13}
\end{equation}

For strong nonrelativistic shocks, we then have
\begin{equation}
\frac{\alpha}{\alpha -3}=\frac{\gamma _2\left( \alpha \right) +1}{\gamma _2\left( \alpha \right) -1},\label{e10}
\end{equation}
that can be solved for $ \alpha $ to obtain self-consistent solutions. Note that $ \gamma _2 $ is associated with $ \alpha $ through its definition in Equations (\ref{e31}) and the EOS indicated above. The system then is determined by $ T $, $ \eta $ and $ p_{\text{m}} $, and solutions with $ u_1<0.1c $ can be treated with nonrelativistic approximations.

Clearly, $ \alpha $ changes from nonrelativistic 4 to relativistic $ 7/2 $ as the gas pressure becomes dominated by relativistic particles. A threshold injection rate $ \eta _{\text{th}} $ for such a transition may be derived with $ P_{\text{nr}}=P_{\text{cr}}\left( \eta _{\text{th}}\right) $ and $ \alpha \rightarrow 4 $:
\begin{equation}
\eta _{\text{th}}=\frac{3}{2}\xi _{T}^{2}\frac{\xi _{\text{m}}-1}{\xi _{\text{m}}\ln \xi _{\text{m}}}.\label{e14}
\end{equation}
This is equivalent to $ \eta _{\text{th}}\propto u_1^2 $ as indicated by Equation (\ref{e8}), implying that for slower shocks it is relatively easier (i.e., having a lower value of $ \eta _{\text{th}} $) for the downstream gas to be relativistic with more efficient acceleration, which may explain the hard radio spectra of old SNRs.

Multi-wavelength observations of SNRs indicate that \citep{2012SSRv..173..369H}
\begin{equation}
T\sim\text{keV},\ cp_\text{m}\gtrsim 10\ \text{GeV}.\label{e24}
\end{equation}
These typical values lead to $ \eta _{\text{th}}\sim 10^{-6} $ as indicated by Equation (\ref{e14}), i.e., a threshold of $ \eta \gtrsim 10^{-6} $ for the downstream gas to be relativistic.

Based on the typical values given by Equation (\ref{e24}), we plot the dependence of $ \alpha $, $ u $ and $ P $ on $ \eta $ in Figure \ref{f2}. The left panel clearly shows that the threshold $ \eta _{\text{th}}\sim 10^{-6} $ for obvious relativistic effects in the downstream. The middle panel shows that $ \eta _{\text{th}} $ tends to increase with $ u_1 $. The right panel shows that the downstream transition from nonrelativistic to relativistic indeed occurs near $ P_{\text{nr}}=P_{\text{cr}} $.

\begin{figure*}
	\includegraphics[width=1\textwidth]{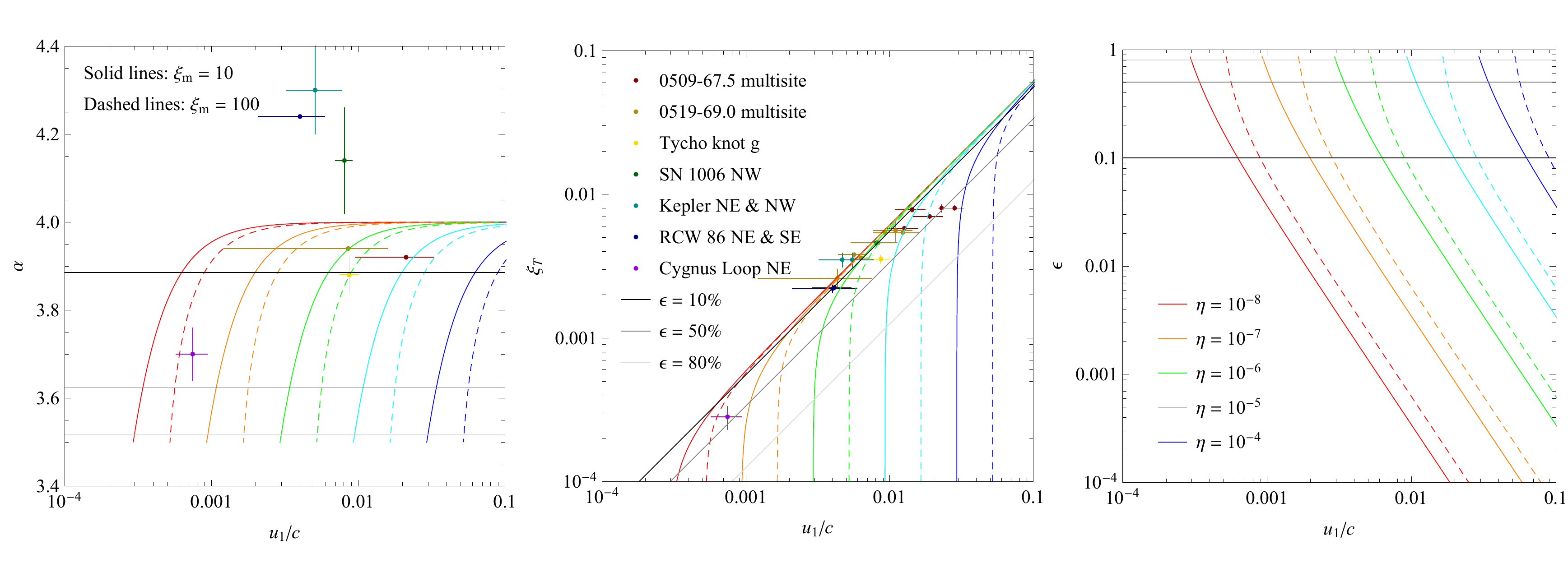}
	\caption{The contours of $ \eta $ with the contour level indicated in the right panel. The CR acceleration efficiency $ \epsilon $ is defined by Equation (\ref{e34}) with $ M\rightarrow \infty $, and the observational data are given in Table \ref{t1}.\label{f3}}
\end{figure*}

The physical processes determining $ \eta $ are likely to be complicated. In Figure \ref{f3}, we plot contours of $ \eta $ in different parameter spaces. The left panel shows again that slower shock tends to have a relativistic downstream gas. It is interesting to note that for given $ \eta $ and $ p_{\text{m}} $ there exits a minimal shock speed $ u_0 $ that can maintain the conservation laws (see also the middle panel). Taking $ P_2=P_{\text{cr}} $ and $ \alpha =7/2 $ in Equation (\ref{e8}), we obtain
\begin{table*}
	\centering
	\caption{Observational data of BDSs}\label{t1}
	\begin{tabular}{l@{\quad}l@{\quad}l@{\quad}l@{\quad}l@{\quad}l@{\quad}l@{\quad}l}
		\hline
		SNR shock & $ \nu $ & $ v_\text{f}/\left( \text{km}/\text{s}\right) $ & $ \omega /\left( \text{arcsec}/\text{yr}\right) $ & $ D/\text{kpc} $ & $ \alpha $ & $ T/\text{eV} $ & $ u_1/\left( \text{km}/\text{s}\right) $\\
		\hline
		0509-67.5 multisite\refn{n12} & 0.46 (F)\refn{n13} & $ 2859\sim 4235 $ & $ 0.012\sim 0.043 $ & $ 50\pm 2 $ & 3.92 & $ \left( 1.54\sim 3.37\right) \times 10^4 $ & $ 2894\sim 9840 $\\
		0519-69.0 multisite\refn{n12} & 0.47 (F)\refn{n13} & $ 1100\sim 3261 $ & $ 0.003\sim 0.021 $ & $ 50\pm 2 $ & 3.94 & $ \left( 0.23\sim 2.00\right) \times 10^4 $ & $ 365\sim 4770 $\\
		Tycho knot g & $ 0.44\pm 0.02 $\refn{n14} & $ 1765\pm 110$\refn{n4} & $ 0.20\pm 0.01 $\refn{n1} & $ 2.5\sim 3.0$\refn{n9} & $ 3.88\pm 0.04 $ & $ 5855^{+753}_{-707} $ & $ 2608^{+379}_{-356} $\\
		SN 1006 NW & $ 0.57\pm 0.06 $ (F)\refn{n15} & $ 2290\pm 80$\refn{n6} & $ 0.30\pm 0.04$\refn{n6} & $ 1.7$\refn{n5} & $ 4.14\pm 0.12 $ & $ 9856^{+701}_{-677} $ & $ 2418\pm 322 $\\
		Kepler NE & $ 0.65\pm 0.05 $ (N)\refn{n16} & $ 1750\pm 200$\refn{n11} & $ 0.060\sim 0.061$\refn{n11} & $ 3.4\sim 6.4$\refn{n3} & $ 4.30\pm 0.10 $ & $ 5756^{+1391}_{-1240} $ & $ 1406^{+445}_{-438} $\\
		Kepler NW & $ 0.65\pm 0.05 $ (N)\refn{n16} & $ 1750\pm 200$\refn{n11} & $ 0.066\sim 0.076$\refn{n11} & $ 3.4\sim 6.4$\refn{n3} & $ 4.30\pm 0.10 $ & $ 5756^{+1391}_{-1240} $ & $ 1650^{+657}_{-586} $\\
		RCW 86 NE & 0.62 (F)\refn{n17} & $ 1100\pm 63$\refn{n8} & $ 0.31\pm 0.08$\refn{n10} & 2.5\refn{n10} & $ 4.24 $ & $ 2274^{+268}_{-253} $ & $ 1204\pm 575$\refn{n10}\\
		RCW 86 SE & 0.62 (F)\refn{n17} & $ 1120\pm 40$\refn{n8} & $ 0.33\pm 0.10$\refn{n10} & 2.5\refn{n10} & $ 4.24 $ & $ 2358^{+171}_{-165} $ & $ 1240\pm 374$\refn{n10}\\
		Cygnus Loop NE & $ 0.35\pm 0.03 $\refn{n18} & $ 115\sim 167$\refn{n2} & $ \left( 3.6\pm 0.5\right) /44$\refn{n2} & $ 0.576\pm 0.061$\refn{n7} & $ 3.70\pm 0.06 $ & $ 37^{+15}_{-13} $ & $ 223^{+58}_{-51} $\\
		\hline
	\end{tabular}
	\begin{tablenotes}
		\item $ \nu $: Radio spectral index. $ \alpha =2\nu +3 $.
		\item $ v_\text{f} $: Full width at half maximum of Gaussian broad component of H$ \alpha $ emission. $ T=m\left( v_\text{f}/2\right) ^2/\left( 2\ln 2\right) $.
		\item $ \omega $: Porper motion of H$ \alpha $ filament. $ D $: Distance from earth to the SNR. $ u_1=\omega D $.
		\item N, S, E, W: North, south, east, west. F: Full remnant.
		\item References: \note{n17}\cite{1975AuJPA..37...39C}, \note{n1}\cite{1978ApJ...224..851K}, \note{n16}\cite{1988ApJ...330..254D}, \note{n18}\cite{1990AJ....100.1927G}, \note{n2}\cite{1999AJ....118..942B}, \note{n3}\cite{1999AJ....118..926R}, \note{n14}\cite{2000ApJ...529..453K}, \note{n15}\cite{2001ApJ...558..739A}, \note{n4}\cite{2001ApJ...547..995G}, \note{n13}\cite{2001ApJ...559..903H}, \note{n5}\cite{2002A&A...387.1047D}, \note{n6}\cite{2002ApJ...572..888G}, \note{n7}\cite{2009ApJ...692..335B}, \note{n8}\cite{2011ApJ...737...85H}, \note{n9}\cite{2011ApJ...729L..15T}, \note{n10}\cite{2013MNRAS.435..910H}, \note{n11}\cite{2016ApJ...817...36S}, \note{n12}\cite{2018ApJ...862..148H}.
	\end{tablenotes}
\end{table*}
\begin{equation}
\frac{u_0}{c}=\frac{7}{3}\sqrt{\frac{\eta}{2}\sqrt{\xi _{\text{m}}}}.\label{e15}
\end{equation}
That is to say, slower shocks tend to have less efficient injection and lower relativistic cutoff, otherwise the slowing down shock speed may hit ``the wall of $ u_0 $''.

The middle panel of Figure \ref{f3} can be divided by the inefficient acceleration line $ T=3mu_1^2/16 $ into two parts: the lower right indicates efficient acceleration, while the upper left is forbidden for violation of conservation laws. The $ u_1 $-$ T $ relation deviates significantly from the dividing line as the acceleration becomes efficient, and $ u_1 $ is independent of $ T $ for extremely efficient acceleration since $ P_\text{cr} $ does not depend on $ T $ for given $ \eta $ and $ p_\text{m} $. Contours of $ \epsilon $ (defined by Equation (\ref{e34}) with $ M\rightarrow \infty $) are also shown in this figure, and are independent of $ p_\text{m} $:
\begin{equation}
\xi _T=\frac{\sqrt{6\left( 2\alpha -7 \right)}}{\alpha}\frac{u_1}{c},
\end{equation}
where $ \alpha $ is associated with $ \epsilon $ via Eqaution (\ref{e33}).

The right panel of Figure \ref{f3} shows dependence of $ \epsilon $ on $ u_1 $, which again implies more efficient acceleration of slower shocks. For linear DSA with $ \alpha \rightarrow 4 $, one has
\begin{equation}
\epsilon =\frac{4}{3}\eta \left( \frac{c}{u_1} \right) ^2\frac{\xi _{\text{m}}\ln \xi _{\text{m}}}{\xi _{\text{m}}-1},
\end{equation}
which should be treated as a lower limit to $ \epsilon $ of self-consistent DSA shock as shown in the figure.

Some observations of Balmer-dominated shocks (BDSs) given in Table \ref{t1} are shown in Figure \ref{f3}. Comparing the left and the middle panel, we find that our model is in good agreement with the Cygnus Loop NE region, therefore we expect the corresponding $ \epsilon $ to be $ \left( 35\pm 10\right) \% $ (Equation (\ref{e33})). For 0509-67.5, 0519-69.0 and Tycho knot g, we conservatively estimate a lower limit of $ \epsilon $ of 7\%, 5\% and 10\%, respectively. For other BDSs with $ \alpha >4 $, since Equation (\ref{e33}) gives $ \epsilon <0 $, no meaningful lower limit to $ \epsilon $ can be estimated with this simple model for strong shocks, but one may consider a finite $ M $ (Equation (\ref{e30})), a slightly concave nonlinear DSA spectrum \citep{1992ApJ...399L..75R}, the time-dependent DSA \citep{2017ApJ...844L...3Z}, etc., to address this issue. Other processes which enhance $ r $ or reduce $ \alpha $ as indicated in the last paragraph of Section \ref{s2} may also be important in generalizing the model. More reliable spatially resolved observations for the SNR's radio spectrum are warranted.

Within uncertainties, observational data in the middle panel of Figure \ref{f3} show $ 10\%\lesssim \epsilon \lesssim 50\% $ for all BDSs, which is consistent with the SNR origin of Galactic CRs. Then Figure \ref{f3} shows that to maintain an $ \epsilon $ of $ 10\%\sim 50\% $ in different stage of SNR's evolution, $ \eta $ and/or $ p_{\text{m}} $ need to increase with $ u_1 $ implying less efficient injection and/or lower relativistic cutoff in older SNRs. Note that this is consistent with what we have predicted above via Eqaution (\ref{e15}) without considering the observational data.

Recent high resolution observation of BDS shows that shocks of older SNRs appear to have a much thinner precursor than younger SNRs \citep{2016ApJ...819L..32K}, we therefore expect more concaved spectra in younger SNRs. More efficient injection is needed for young SNRs to accelerate particle efficiently. On the other hand, it is possible that even young SNRs may dominate the acceleration of CRs beyond 100 GeV, the overall acceleration efficiency is much lower than 10\%. Observations of Cas A do indicate a very low acceleration efficiency \citep{2010ApJ...722.1727V}. If the injection rate does not change with the shock speed (above $ u_0 $), we would expect the overall acceleration efficiency increase rapidly as the shock slows down as shown in the right panel of Figure \ref{f3}. However, since observations and theoretical considerations suggest that the maximum energy of the accelerated particles increases with the shock speed, which may shift ``the barrier of $ u_0 $'' to an even lower value, the dependence of overall acceleration efficiency on the shock speed should be more gradual than that in Figure \ref{f3}.

\section{Conclusion and Discussion}\label{s4}
Based on the classical Rankine-Hugoniot shock conditions, we study correlations between the CR acceleration efficiency and the power-law spectral index of the downstream nonthermal particle distribution in momentum space produced via DSA with a shock speed much smaller than the speed of light. The value of the spectral index takes the steady-state DSA value for test particles given by Equation (\ref{e38}) as we are interested in the global characteristics of the shock acceleration and ignore structure of the shock precursor for the sake of simplicity. Our model is compatible with most nonlinear DSA models except that the actual spectral index of accelerated particles should be greater than $ \alpha $ given by Equation (\ref{e38}).

Our results (Equations (\ref{e34}) $ \sim $ (\ref{e36})) show an anticorrelation between the acceleration efficiency and the spectral index, which, in combination with the observed radio spectral hardening of old SNRs, suggests that the acceleration efficiency increases gradually as the shock of SNR slows down. We find that escape of CRs from the particle accelerating shocks is not necessary for efficient particle acceleration, which contradicts some nonlinear DSA models \citep{2010ApJ...722.1727V}, where the effect of accelerated particles on the dynamics of subshock is ignored. Although the spectral index of radio emitting electrons may be affected by many processes, the simplest model proposed here shows that a radio spectral index of 0.44 implies a more than 10\% acceleration efficiency. For the NE region of Cygnus Loop, an efficiency of $ \left( 35\pm 10\right) \% $ is inferred. Detailed modeling of individual SNRs interacting with molecular clouds is necessary to have a better measurement of the acceleration efficiency.

Assuming a Maxwell distribution for nonrelativistic particles plus a power-law distribution with a high-energy cutoff for relativistic particles in the downstream of strong shocks, we find that for typical parameters of SNRs, the relativistic particles injected into the DSA process needs to be more than $ \sim 10^{-6} $ times the nonrelativistic thermal particles for highly efficient particle acceleration and this threshold injection rate increases with the shock speed since it is more difficult for relativistic particles to dominate the downstream pressure of higher speed shocks. Most nonlinear DSA models predict a concave spectral shape, the threshold injection rate for relativistic particles to dominate is even higher in these scenarios. On the other hand, if nonrelativistic particle acceleration is dominated by resonances with plasma waves, a convex spectral shape is expected near the particle rest mass energy \citep{2004ApJ...610..550P} and we expect a lower threshold injection rate for efficient particle acceleration.

Our results also show that the injection rate and/or the relativistic cutoff energy need to be low for old SNRs. If the injection rate does not change with the shock speeds, we expect that the acceleration efficiency increases significantly as SNR evolves with its shock slowing down. This acceleration efficiency increase will be less dramatic than that shown in Figure \ref{f3} if one takes into account the fact that the maximum energy of accelerated particles increases quickly with the shock speed \citep{2004MNRAS.353..550B}, we therefore expect a gradual increase of the overall acceleration efficiency as SNR evolves.

This is consistent with the scenario that Galactic CRs are mostly produced in SNRs, where SNRs not only are powerful enough to sustain the overall CR flux but also can accelerate particles efficiently to the PeV energies. According to our studies, although high-energy Galactic CRs are mostly accelerated in early stage of SNR evolution, the overall acceleration efficiency in these young SNRs can be much less than 10\% for the low flux of high-energy CRs. Most of the low-energy CRs are accelerated in old SNRs with harder radio spectra. The corresponding acceleration efficiency should be greater than 10\%, consistent with an SNR origin of Galactic CRs \citep{2017ApJ...844L...3Z}. Our results therefore suggest less efficient particle acceleration in younger SNRs with a relatively softer spectrum cutting off at higher energies. The bulk of Galactic CR acceleration occurs in old SNRs with harder spectra and lower cutoff energies. The model can not only explain the gradual hardening of SNR's radio spectrum with age, but also account for the convex $ \gamma $-ray spectrum and the spectral knee of CRs \citep{2011ApJ...729L..13O}. Recent observations of BDSs also support such a scenario \citep{2016ApJ...819L..32K}. More detailed modeling is warranted.

The thermal equilibrium between proton and electron may also not be ignored in general. The temperature $ T $ in Section \ref{s3} must be interpreted as the proton temperature, thus such proton-electron thermal equilibrium clearly makes $ T $ lower for given shock speed and maximum momentum than that in Section \ref{s3}, or roughly makes the solutions shown in the middle panel of Figure \ref{f3} shift downwards slightly, because some of the thermal energies are shared by electrons. The shock geometry and effects of large-scale magnetic fields also need to be considered when applying this model to observations of individual SNRs.

The method of this work can be extended to the relativistic DSA shocks as long as we know exact fomula of the particle spectral index in terms of relativistic up- and downstream fluid speeds \citep{2005PhRvL..94k1102K}. With such scenario, we may constrain the DSA in more extreme astrophysical environments, such as pulsar wind nebulae and so on.
\section*{Acknowledgements}
This work is partially supported by National Key Research \& Development Program of China: 2018YFA0404203, National Natural Science Foundation of China grants: U1738122, 11761131007 and by the International Partnership Program of Chinese Academy of Sciences, grant no. 114332KYSB20170008.
\bibliographystyle{mnras}

\end{document}